\documentclass{jfm}
\usepackage{amsmath}
\usepackage{amssymb}
\usepackage{apacite}
\usepackage{graphicx}
\usepackage{epstopdf, epsfig}
\usepackage{floatrow}
\usepackage{subfig}
\usepackage[dvipsnames]{xcolor}

\begin{document}
\newtheorem{lemma}{Lemma}
\newtheorem{corollary}{Corollary}
\shorttitle{Velocity from scalar field} 
\shortauthor{Sharma et al.} 
\title{Analytic Reconstruction of a Two-Dimensional Velocity Field from an Observed Diffusive Scalar}
\author
{
	Arjun Sharma\aff{1},
	Irina I. Rypina \aff{2},
	Ruth Musgrave\aff{2}
	\and
George Haller\aff{1}
	\corresp{\email{georgehaller@ethz.ch}}
}
\affiliation
{	\aff{1}
	Institute for Mechanical Systems, ETH Zurich, Leonhardstrasse 21, 8092 Zurich, Switzerland\\
	\aff{2}
	Woods Hole Oceanographic Institution, 266 Woods Hole rd., Woods Hole, MA 02543, USA
}
\maketitle
\begin{abstract}
Inverting an evolving diffusive scalar field to reconstruct the underlying velocity field is an underdetermined problem. Here we show, however, that for two-dimensional incompressible flows, this inverse problem can still be uniquely solved if high-resolution tracer measurements, as well as velocity measurements along a curve transverse to the instantaneous scalar contours, are available. Such measurements enable solving a system of partial differential equations for the velocity components by the method of characteristics. If the value of the scalar diffusivity is known, then knowledge of just one velocity component along a transverse initial curve is sufficient. These conclusions extend to the shallow-water equations and to flows with spatially dependent diffusivity. We illustrate our results on velocity reconstruction from tracer fields for planar Navier\textendash Stokes flows and for a barotropic ocean circulation model. {We also discuss the use of the proposed velocity reconstruction in oceanographic applications to extend localised velocity measurements to larger spatial domains with the help of remotely sensed scalar fields.}
\end{abstract}
\section{Introduction and related work}\label{sec:Introduction}
The measurement of an evolving scalar tracer is sometimes considered easier than that of the underlying velocity \citep{schlichtholz1991review}, both in nature and in a laboratory setting. Determining the flow velocity more accurately from scalar field observations is, therefore, an exciting perspective in a plethora of applications including oceanography, meteorology, medicine and chemical engineering \citep{wunsch1987using}.

The transport equation for a passive scalar with instantaneous concentration
$c(\mathbf{x},t)$ and constant diffusivity $\mu\geq0$ under an incompressible,
planar velocity field $\mathbf{u}(\mathbf{x},t)$ is given by
\begin{equation}
\frac{\partial c}{\partial t}+\mathbf{u}\cdot\mathbf{\nabla}c=\mu\nabla^{2}c,\label{eq:Transport Equation}
\end{equation}
with the spatial coordinates $\mathbf{x}=(x,y)$ varying over some
domain $U\subset\mathbb{R}^{2}$ and $t$ denoting the time. Note
that the velocity field $\mathbf{u}$ only influences the advection of $c$ via its inner product $\mathbf{u}\cdot\mathbf{\nabla}c$
with the instantaneous concentration gradient. This leads to the classic
observation that the iso-scalar component of the velocity is inconsequential
for the scalar field evolution \citep{dahm1992scalar,kelly1989inverse}.
In other words, the iso-scalar velocity component falls in the null
space of the velocity-reconstruction problem, necessitating further
conditions for a unique solution to this inverse problem \citep{kelly1989inverse}.

Seminal work by \citet{fiadeiro1984obtaining} and \citet{wunsch1985can} considered velocity reconstruction from discretized observations of a steady scalar distribution in a two-dimensional, steady channel flow with uniform velocity. After spatial discretization, these authors obtained an under-determined algebraic system of equations for the values of the streamfunction at the grid points. The minimum-norm solution obtained to this problem by \citet{fiadeiro1984obtaining} and \citet{wunsch1985can} was found to be qualitatively inaccurate by \citet{veronis1986comments} even for the simple uniform flow considered. Providing additional boundary conditions \citep{fiadeiro1984obtaining} or information on the flow anisotropy \citep{wunsch1985can} leads to improved solutions. Alternatively, if the flow is dispensed with two scalars of different diffusivity, the system becomes over-determined because the same streamfunction must satisfy two linearly independent scalar equations. This reformulation, however, allows for a physically realistic solution that minimizes the least-squares of the solution error, as obtained by \citet{fiadeiro1984obtaining}. As expected, the reconstructed velocity departs from the true velocity significantly in regions where the two scalar fields are similarly distributed.

An alternative approach to inverting the scalar transport equation \eqref{eq:Transport Equation} has led to the emergence of Scalar Imaging Velocimetry (SIV) (see \citealt{wallace2010measurement} for a review). When the diffusion of the scalar is slower than that of the carrier fluid (i.e., the ratio of fluid viscosity and scalar diffusivities is larger than one), the fluid velocity at nearby points varies slower than the scalar intensity \citep{dahm1992scalar}. As a consequence, the velocity component normal to the iso-scalar varies faster than the overall velocity. For an initial estimate, the overall velocity can therefore be assumed locally spatially constant and recovered from a projection onto neighboring normals to iso-scalar curves. As the calculation of iso-scalar-normal velocity requires division by $|\nabla c|$, the velocity reconstructed in this fashion is limited to regions where $|\nabla c|$ is well separated from zero \citep{dahm1992scalar}.

In an another approach relaxing the requirement of slow scalar diffusion, \citet{su1996scalar} formulated the velocity reconstruction as a minimization problem for an integral composed of the transport equation, the divergence and a regularizer term depending on the velocity gradient norm. \citet{kelly1989inverse} also implemented a similar formulation for ocean flows in which the functional to be minimized involves the transport equation, the horizontal divergence, the relative vorticity and the kinetic energy. These minimization techniques can return an accurate (albeit never exact) velocity field if the weight factors for each component of the functional are tuned appropriately.

Here, we point out that for two-dimensional incompressible flows, one can use the method of characteristics to analytically reformulate the partial differential equation (PDE) \eqref{eq:Transport Equation} into a system of ordinary differential equations (ODEs) for the spatial evolution of the velocity components along an instantaneous level curve of $c(\mathbf{x},t)$. Integration of these ODEs enables the reconstruction of the velocity field over the domain of scalar observations. This procedure assumes that the velocity is available along a curve $\Gamma(t)$ transverse to the level curves of $c(\mathbf{x},t)$. If the value of the diffusivity $\mu$ is known, knowledge of one velocity component along $\Gamma(t)$ turns out to be sufficient. We give the derivation of these results in section \ref{sec:Formulation} and test the approach on two numerical examples in section \ref{sec:Numerics}. 

Our technique can be particularly advantageous in oceanographic applications, in which often only localised velocity field measurements are available. In such situations, independent observations of larger-scale physical, chemical or biological tracer fields offer an opportunity to extend the local velocity data to larger domains, as we illustrate in section \ref{sec:Oceanographic-applications}. We also discuss oceanography-inspired extensions of our main results to diffusivities with spatial dependence and to shallow-water equations in section \ref{sec:Formulation}.

\section{Formulation}

\label{sec:Formulation} Taking the spatial gradient of the scalar transport equation \eqref{eq:Transport Equation} leads to the equations
\begin{subequations} 
	\begin{eqnarray}
	\frac{\partial^{2}c}{\partial t\partial x}+u\frac{\partial^{2}c}{\partial x^{2}}+\frac{\partial u}{\partial x}\frac{\partial c}{\partial x}+v\frac{\partial^{2}c}{\partial x\partial y}+\frac{\partial v}{\partial x}\frac{\partial c}{\partial y}=\mu\frac{\partial{\nabla^{2}{c}}}{\partial x},\\
	\frac{\partial^{2}c}{\partial t\partial y}+u\frac{\partial^{2}c}{\partial x\partial y}+\frac{\partial u}{\partial y}\frac{\partial c}{\partial x}+v\frac{\partial^{2}c}{\partial y^{2}}+\frac{\partial v}{\partial y}\frac{\partial c}{\partial y}=\mu\frac{\partial{\nabla^{2}{c}}}{\partial y}.
	\end{eqnarray}
	\label{eq:transpt1} \end{subequations} 
There is no general recipe for solving such a coupled system of linear PDEs. Remarkably, however, the incompressibility condition
\begin{equation}
\frac{\partial u}{\partial x}+\frac{\partial v}{\partial y}=0\label{eq:incomp}
\end{equation}
enables us to rewrite this system as \begin{subequations} 
	\begin{align}
	\frac{\partial v}{\partial x}\frac{\partial c}{\partial y}-\frac{\partial v}{\partial y}\frac{\partial c}{\partial x}=-u\frac{\partial^{2}c}{\partial x^{2}}-v\frac{\partial^{2}c}{\partial x\partial y}-\frac{\partial^{2}c}{\partial t\partial x}+\mu\frac{\partial{\nabla^{2}{c}}}{\partial x},\\
	\frac{\partial u}{\partial x}\frac{\partial c}{\partial y}-\frac{\partial u}{\partial y}\frac{\partial c}{\partial x}=\hspace{0.15in}u\frac{\partial^{2}c}{\partial x\partial y}+v\frac{\partial^{2}c}{\partial y^{2}}+\frac{\partial^{2}c}{\partial t\partial y}-\mu\frac{\partial{\nabla^{2}{c}}}{\partial y}.
	\end{align}
	\label{eq:transpt2} \end{subequations}Both equations in system \eqref{eq:transpt2}
share the same characteristics $\mathbf{x}(s;t)=(x(s;t),y(s;t))$, which satisfy the system of ODEs 
\begin{equation}
\frac{dx(s;t)}{ds}=\frac{\partial c(\mathbf{x}(s;t),t)}{\partial y},\quad\frac{dy(s;t)}{ds}=-\frac{\partial c(\mathbf{x}(s;t),t)}{\partial x},\label{eq:characteristics}
\end{equation}
with the instantaneous time $t$ playing the role of a parameter. These characteristics are pointwise normal to the gradient $\nabla c$ and hence coincide with instantaneous level curves of $c(\mathbf{x},t)$ at any time $t$. Along the characteristics \eqref{eq:characteristics}, the PDEs in \eqref{eq:transpt2} can be re-written as a linear, two-dimensional, non-autonomous system of ODEs for the velocity $\mathbf{\tilde{u}}(s;t)\mathrel{\mathop{:}}=\mathbf{u}(\mathbf{x}(s;t),t)$. This ODE is of the form 
\begin{equation}
\frac{d\mathbf{\mathbf{\tilde{u}}}(s;t)}{ds}=\mathbf{A}(s;t)\mathbf{\tilde{u}}(s;t)+\mathbf{b}(s;t),\quad\mathbf{\tilde{u}}(s;t)\in\mathbb{R}^{2},\quad s\in\mathbb{R},\label{eq:ode1}
\end{equation}
with the form of the matrix $\mathbf{A}(s;t)\in\mathbb{R}^{2\times2}$ and the vector $\mathbf{b}(s;t)\in\mathbb{R}^{2}$ depending upon wether the constant scalar diffusivity, $\mu$, is known or unknown, as we discuss below separately. In either case, once the normalized fundamental matrix solution $\boldsymbol{\Phi}(s;t)$ of the homogenous part $\frac{d\mathbf{\tilde{u}}}{ds}=\mathbf{A}(s;t)\mathbf{\tilde{u}}$, satisfying $\boldsymbol{\Phi}(0;t)=\mathbf{I}$, has been numerically determined, we can integrate \eqref{eq:ode1} directly to obtain the velocity
\begin{equation}
\mathbf{\tilde{u}}(s;t)=\boldsymbol{\Phi}(s;t)\mathbf{\tilde{u}}(0;t)+\int_{0}^{s}\boldsymbol{\Phi}(s-\sigma;t)\mathbf{b}(\sigma;t)d\sigma\label{eq:integralform}
\end{equation}
at any location $\mathbf{x}(s;t)$ along the level curve of $c(\mathbf{x},t)$ containing the point $\mathbf{x}(0;t)$.

\subsection{Case of known scalar diffusivity}

In the case where the value of $\mu$ is \textit{a priori} known, equations \eqref{eq:transpt2} can be recast in the form \eqref{eq:ode1} with the help of 
\begin{equation}
\mathbf{A}(s;t)=\begin{bmatrix}\frac{\partial^{2}c}{\partial x\partial y} & \frac{\partial^{2}c}{\partial y^{2}}\\
-\frac{\partial^{2}c}{\partial x^{2}} & -\frac{\partial^{2}c}{\partial x\partial y}
\end{bmatrix}_{\mathbf{x}=\mathbf{x}(s;t)},\quad\mathbf{b}(s;t)=\begin{bmatrix}\frac{\partial^{2}c}{\partial t\partial y}-\mu\frac{\partial{\nabla^{2}{c}}}{\partial y}\\
-\frac{\partial^{2}c}{\partial t\partial x}+\mu\frac{\partial{\nabla^{2}{c}}}{\partial x}
\end{bmatrix}_{\mathbf{x}=\mathbf{x}(s;t)}.\label{eq:known_diff_form}
\end{equation}

\subsection{Case of unknown scalar diffusivity}\label{sec:unknownDiff}

If the value of $\mu$ is \textit{a priori} unknown, we can still express $\mu$ from the original transport equation \eqref{eq:Transport Equation} as 
\begin{equation}
\mu=\frac{1}{\nabla^{2}{c}}\Big(\frac{\partial c}{\partial t}+u\frac{\partial c}{\partial x}+v\frac{\partial c}{\partial y}\Big).\label{eq:diffusivity}
\end{equation}
Substitution of this expression into \eqref{eq:transpt2} enables us to rewrite \eqref{eq:transpt2} in the form \eqref{eq:ode1} with
\begin{align}
\begin{split}\mathbf{A}(s;t) & =\begin{bmatrix}\frac{\partial^{2}c}{\partial x\partial y}-\frac{1}{\nabla^{2}{c}}\frac{\partial c}{\partial x}\frac{\partial{\nabla^{2}c}}{\partial y} & \frac{\partial^{2}c}{\partial y^{2}}-\frac{1}{\nabla^{2}{c}}\frac{\partial c}{\partial y}\frac{\partial{\nabla^{2}{c}}}{\partial y}\ \\
-\frac{\partial^{2}c}{\partial x^{2}}+\frac{1}{\nabla^{2}{c}}\frac{\partial c}{\partial x}\frac{\partial{\nabla^{2}{c}}}{\partial x} & -\frac{\partial^{2}c}{\partial x\partial y}+\frac{1}{\nabla^{2}{c}}\frac{\partial c}{\partial y}\frac{\partial{\nabla^{2}{c}}}{\partial x}
\end{bmatrix}_{\mathbf{x}=\mathbf{x}(s;t)},\\
\quad\mathbf{b}(s;t) & =\begin{bmatrix}\frac{\partial^{2}c}{\partial t\partial y}-\frac{1}{\nabla^{2}{c}}\frac{\partial c}{\partial t}\frac{\partial{\nabla^{2}{c}}}{\partial y}\\
-\frac{\partial^{2}c}{\partial t\partial x}+\frac{1}{\nabla^{2}{c}}\frac{\partial c}{\partial t}\frac{\partial{\nabla^{2}{c}}}{\partial x}
\end{bmatrix}_{\mathbf{x}=\mathbf{x}(s;t)}.
\end{split}
\label{eq:unknown_diff_form}
\end{align}
The formulae in equation \eqref{eq:unknown_diff_form}, however, are only well-defined at points where $\nabla^{2}c\ne\mathbf{0}$.

\subsection{Spatially-dependent diffusivity\label{subsec:Spatially-dependent-diffusivity}}

Most non-eddy-resolving ocean circulation models are diffusion-based, i.e., assume that the small-scale, un- and under-resolved eddy motions lead to a diffusive tracer transfer\textemdash a process that can be fully characterized by just one parameter, the lateral eddy diffusivity $\mu $. A number of diffusivity estimates are available based on data from surface drifters and tracers \citep{okubo1971oceanic, zhurbas2004drifter, lacasce2008statistics, rypina2012eddy, lacasce2000relative, lumpkin2002lagrangian, mcclean2002eulerian} satellite-observed velocity fields \citep{marshall2006estimates,abernathey2013global,klocker2014global,rypina2012eddy} and Argo float observations \citep{cole2015eddy}. All these observations suggest that the ocean eddy diffusivity is strongly spatially varying with up to two orders of magnitude difference between regions with quiescent vs. energetic oceanic eddy regimes. Eddy-resolving simulations also demonstrate highly nonuniform spatial distribution of the eddy-induced transport \citep{gille1999influence,nakamura2000eddy,roberts2000validity}. Thus, any method that involves diffusivity assumption in a global settings, including the velocity reconstruction method described here, would need to take into account spatial variation in diffusivity.

In this case, the appropriate advection-diffusion equation takes the form 
\begin{equation}
\frac{\partial c}{\partial t}+\mathbf{u}\cdot\mathbf{\nabla}c=\nabla(\mu \nabla c).\label{eq:adkappa}
\end{equation}
Taking the gradient of this equation and following the same steps as in Section 2, one obtains that the velocity $\tilde{\mathbf{u}}$ satisfies equation \eqref{eq:ode1} along the iso-curves of tracer concentration $c$, with $\mathbf{A}$ from eq. \eqref{eq:known_diff_form} and 
\begin{equation}\label{eq:varying_diff_forcing}
\mathbf{b}(s;t)=\begin{bmatrix}\frac{\partial^{2}c}{\partial t\partial y}-\frac{\partial{\nabla(\mu \nabla{c})}}{\partial y}\\
-\frac{\partial^{2}c}{\partial t\partial x}+\frac{\partial{\nabla(\mu \nabla{c})}}{\partial x}
\end{bmatrix}_{\mathbf{x}=\mathbf{x}(s;t)}.
\end{equation}

\subsection{Shallow-water approximation\label{subsec:Shallow-water-approximation}}

Due to negligible vertical velocities, eq. \eqref{eq:incomp} is a reasonable approximation for oceanic surface flows on spatial scales larger than the mesoscale \citep{pedlosky2013geophysical,salmon1998lectures,bennett2005inverse}. When vertical velocities become non-negligible, one may still find another variable other than the velocity that satisfies eq. \eqref{eq:incomp} and re-derive the inversion for that variable. An example is the inviscid shallow-water model with a rigid lid \citep{pedlosky2013geophysical,batchelor1967introduction}, as we shall discuss next. This shallow-water approximation is often employed in physical oceanography to describe ocean flows in shallow tidally-driven bays, such as, for example, the Katama Bay located on the island of Martha's Vineyard, MA \citep{slivinski2017assimilating,orescanin2016changes,luettich1991solution}.

Consider a shallow sheet of inviscid unstratified fluid overlaying spatially-dependent bathymetry. Fluid parcels are allowed to have three non-zero components of velocity, $\{u,v,w\}\neq0$, but if the horizontal velocity $\mathbf{u}$ is initially independent of $z$ then it remains so for all times, i.e., we may write
\begin{equation}
\mathbf{u}=\mathbf{u}(x,y,t).
\end{equation}
If the initial tracer distribution is independent of $z$, i.e., $c_{0}=c_{0}(x,y)$, then we will also have
\begin{equation}
c=c(x,y,t)
\end{equation}
for all times. We then invoke a rigid-lid approximation, assuming that the free surface elevation $\eta$\textemdash the deviation of the free surface from its mean value\textemdash is much smaller than the water depth $h(x,y)$, i.e., $\eta\ll h\ll L$, where $L$ is the characteristic horizontal length scale of motion. Then the water depth is time-independent and can be considered known because the bathymetry of a shallow bay can be mapped out and the sea surface height is assumed to be constant due to the rigid lid approximation. 

In this case, mass conservation requires the transport velocity, $h\mathbf{u}$, to be divergence free:
\begin{equation}
\partial(hu)/\partial x+\partial(hv)/\partial y=0.
\end{equation}
To proceed, we multiply the advection- diffusion equation \eqref{eq:adkappa} by $h$ to obtain
\begin{equation}
h\frac{\partial c}{\partial t}+h\mathbf{u}\cdot\mathbf{\nabla}c=h\nabla(\mu \nabla c).\label{eq:new PDE}
\end{equation}
We take the gradient of this equation, then follow the steps of section 2 to obtain an analogue of eq. \eqref{eq:ode1} for the transport velocity, given by
\begin{equation}
d(h\tilde{\mathbf{u}})/ds=\mathbf{A}(h\tilde{\mathbf{u}})+\mathbf{b},\label{eq:transvel}
\end{equation}
with $\mathbf{A}$ from eq. \eqref{eq:known_diff_form} and with
\begin{equation}\label{eq:forcing_shallow_water}
\mathbf{b}(s;t)=\begin{bmatrix}\frac{\partial}{\partial y}(h\frac{\partial c}{\partial y})-\frac{\partial}{\partial y}(h\nabla(\mu \nabla{c}))\\
-\frac{\partial}{\partial x}(h\frac{\partial c}{\partial y})+\frac{\partial}{\partial x}(h\nabla(\mu \nabla{c}))
\end{bmatrix}_{\mathbf{x}=\mathbf{x}(s;t)}.
\end{equation}
As before, equation \eqref{eq:transvel} should be solved along the iso-contours of $c$, which are the characteristics of the  PDE \eqref{eq:new PDE}.

\subsection{Velocity reconstruction}

\label{subsec:procedure_velocity} For an experimentally or numerically obtained scalar field, $c(\mathbf{x},t)$, the characteristic $\mathbf{x}(s;t)$ emanating from any point $\mathbf{x}_{0}(t)\mathrel{\mathop{:}}=\mathbf{x}(0;t)$ can be computed numerically from equation \eqref{eq:characteristics}. To recover the velocity $\mathbf{\tilde{u}}(s;t)$ at a point $\mathbf{x}(s;t)$ of such a characteristic via \eqref{eq:integralform}, we need to know the velocity $\mathbf{\tilde{u}}(0;t)$ at the initial location $\mathbf{x}_{0}(t)$. The PDE \eqref{eq:transpt2} is well-posed only if such a set of initial velocity vectors is available along a non-characteristic curve $\Gamma(t)$, i.e., along a curve segment that is transverse to the characteristics at each of its points. If, however, the diffusivity $\mu$ is a priori available, then only one component of $\mathbf{\tilde{u}}(0;t)=\mathbf{{u}}(\mathbf{x}_{0},t)$ needs to be known along the non-characteristic curve $\Gamma(t)$, because the other component can be determined from \eqref{eq:Transport Equation}. For example, if only the $u(\mathbf{x},t)$ velocity is known at a point of $\mathbf{x}\in\Gamma(t)$, then $v(\mathbf{x},t)$ can be uniquely determined from \eqref{eq:Transport Equation} as long as the contour of $c(\mathbf{x},t)$ is not locally parallel to the $x$ direction, i.e., as long as ${\partial c}(\mathbf{x},t)/{\partial y}\ne0$.

To reconstruct the velocity field $\mathbf{u}(\mathbf{x},t)$ over the largest possible domain, the initial velocities must therefore be known along a curve $\Gamma(t)$ transverse to the level curves of $c(\mathbf{x},t)$ over the largest possible domain. Most of the characteristics (level curves) in our examples will be closed, but this is not a requirement for our method to be applicable. Indeed, to obtain $\mathbf{u}(\mathbf{x},t)$ along a non-closed level set of $c(\mathbf{x},t)$, one simply has to integrate equation \eqref{eq:characteristics} both in the forward and the backward $s$-direction until the domain boundary is reached.

The formulae \eqref{eq:known_diff_form}-\eqref{eq:forcing_shallow_water} contain the first-order temporal derivative of the observed scalar $c(\mathbf{x},t)$, as well as up to second and up to third-order spatial derivatives, respectively. This presents challenges for velocity reconstruction over domains where only scarce information is available on $c(\mathbf{x},t)$. We envision, however, applications of this velocity reconstruction procedure in situations where high-resolution imaging of $c(\mathbf{x},t)$ is available, enabling an accurate computation of the necessary derivatives.

{In geophysical flows, the ratio of advective transport rate to the diffusive rate (i.e. the dimensionless P\'eclet number) tends to be large. For such large P\'eclet numbers, in the formulae relevant for velocity reconstruction under a known diffusivity (eqs. \eqref{eq:known_diff_form}, \eqref{eq:unknown_diff_form}, \eqref{eq:varying_diff_forcing} and \eqref{eq:forcing_shallow_water}), errors and uncertainties arising from computing the necessary third derivatives are suppressed by the small $\mu$ values multiplying these terms. We illustrate this effect in section \ref{sec:NoiseandUndersample} with respect to substantial under-sampling and moderate uncertainties.}

\section{Numerical examples} \label{sec:Numerics} 
We now carry out our proposed velocity reconstruction scheme in two examples. In both, we generate $c(\mathbf{x},t)$ numerically by evolving an initial scalar distribution $c_{0}(\mathbf{x})\mathrel{\mathop{:}}=c(\mathbf{x},t_{0})$ under the PDE \eqref{eq:Transport Equation} with a known velocity field $\mathbf{u}(\mathbf{x},t)$, then reconstruct $\mathbf{u}(\mathbf{x},t)$ solely from observations of $c(\mathbf{x},t)$. We solve for $c(\mathbf{x},t)$ by discretizing \eqref{eq:Transport Equation} using the pseudospectral method. The necessary spatial derivatives of $c(\mathbf{x},t)$ in the reconstruction formulae are evaluated in the frequency domain using MATLAB's FFT algorithm; the time integration is performed in real space using MATLAB's variable-order adaptive time step solver, ODE45. The standard 2/3 dealiasing is applied where the energy in the last one third of the wavenumbers is artificially removed. In both examples, the scalar diffusivity is taken to be $\mu=0.01$ and a 512 $\times$ 512 grid is used. We perform the integration of the characteristic ODEs (equations \eqref{eq:characteristics} and \eqref{eq:ode1}) using MATLAB's ODE113 to achieve high accuracy.

\subsection{Example 1: Steady, analytic, inviscid velocity field}

\begin{figure}
	\centering
	\includegraphics[width=1\textwidth]{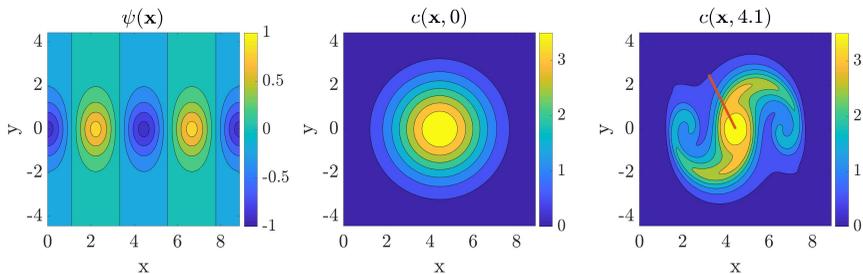} 
	\caption {Contour plot of the steady-state streamfunction for Mallier\textendash Maslowe vortices with $\alpha=1$ in equation \eqref{eq:Maslowe Vortices} (left) and the evolution of the initial concentration field (middle) along these vortices into the one at $t=4.1$ (right) for the scalar diffusivity value $\mu=0.01$.  \label{fig:Maslowe Vortices}}
\end{figure}

We consider the Mallier\textendash Maslowe vortices \citep{mallier1993row,gurarie2004vortex} which are exact solutions of the 2D Euler equations. Modeling a periodic row of vortices with alternating rotation directions, the corresponding velocity field derives from the streamfunction
\begin{equation}
\psi(\mathbf{x})=\text{arctanh}\Big[\frac{\alpha\cos\sqrt{1+\alpha^{2}}x}{\sqrt{1+\alpha^{2}}\cosh{\alpha y}}\Big],\label{eq:Maslowe Vortices}
\end{equation}
which we plot for reference in the left panel of figure \ref{fig:Maslowe Vortices} for $\alpha=1$. 
\floatsetup[figure]{style=plain,subcapbesideposition=top}

Over the square domain $\mathbf{x}\in[0,2\sqrt{2}\pi]\times[-\sqrt{2}\pi,\sqrt{2}\pi]$, we generate the diffusive scalar distribution $c(\mathbf{x},t)$ from the initial concentration field $c(\mathbf{x},0)=4\exp[-0.2((x-\sqrt{2}\pi)^{2}+y^{2})]$ by solving the transport equation \eqref{eq:Transport Equation}. No advective transport is possible among the cells defined by the streamfunction in the left panel of figure \ref{fig:Maslowe Vortices}, but diffusion enables the transport of the scalar $c$ across cell boundaries. We will reconstruct the underlying velocity field generated by \eqref{eq:Maslowe Vortices} at the time $t=4.1$. The initial and the evolved scalar fields are shown in the middle and right panels of figure \ref{fig:Maslowe Vortices}. The non-characteristic initial curve $\Gamma(t)$ for the characteristics is formed by a set of 100 points on the solid red curve in the right panel of figure \ref{fig:Maslowe Vortices}, along which we consider the velocities known.

\begin{figure}
	\centering
	\sidesubfloat[$u(\mathbf{x},4.1)$]{\includegraphics[width=1\textwidth]{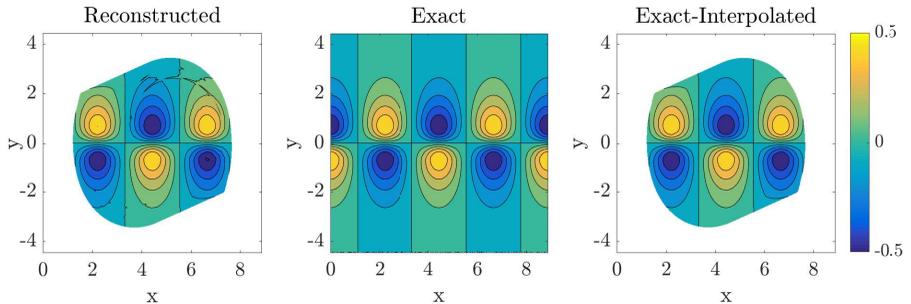}\label{fig:u_vel_analyt} }\\
	\sidesubfloat[$v(\mathbf{x},4.1)$]{\includegraphics[width=1\textwidth]{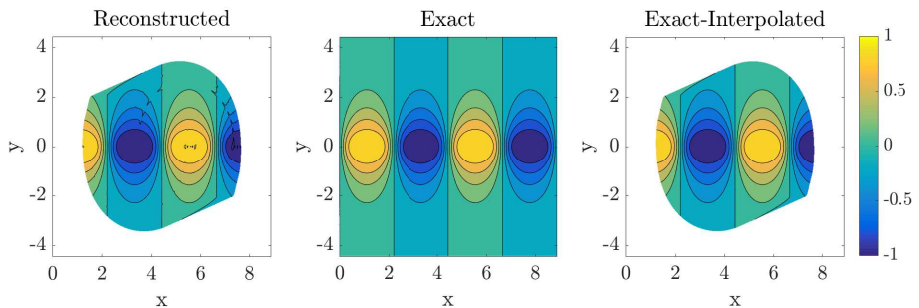} \label{fig:v_vel_analyt}}
	\caption {Reconstructed, exact and exact-interpolated velocities for Example 1 at $t=4.1$, with the diffusivity assumed unknown: (a) $u(\mathbf{x},4.1)$ (b) $v(\mathbf{x},4.1)$. The reconstructed velocities have been interpolated from the characteristics (level sets in the right panel of figure \ref{fig:Maslowe Vortices}) to a uniform grid. The exact-interpolated velocities are obtained by evaluating equation \eqref{eq:Maslowe Vortices} over the characteristics at $t=4.1$, then interpolating to the same uniform grid used for the reconstructed velocities. This procedure is designed to factor out interpolation errors that would be present in a direct comparison with the exact velocity field evaluated directly over the numerical grid. \label{fig:vel_comvel_comp_uniform_unkn_diff}}
\end{figure}

Next, we reconstruct the velocity field using both the known- and
the unknown-diffusivity formulation. While integrating equation \eqref{eq:Transport Equation},
we record velocities along each characteristic at 3000 locations,
irrespective of their lengths. We define the relative normed error
over the computational domain as the ratio of the $L_{2}$ norm of
the difference between the reconstructed and the original velocity
field to the $L_{2}$ norm of the original velocity field defined
in equation \eqref{eq:Maslowe Vortices}. For the ${u}$ velocity
component, this error norm, $E_{u}$, can be written as 
\begin{equation}
E_{u}=\frac{||u_{\text{reconst.}}-u_{\text{exact}}||_{2}}{||u_{\text{exact}}||_{2}},\label{eq:normed_error}
\end{equation}
where $u_{\text{reconst.}}$ and $u_{\text{exact}}$ are the reconstructed
and exact $u$-velocities, respectively. The error, $E_{v}$, for
the $v$ velocity component is defined similarly. We have found $E_{u}$
and $E_{v}$ for the known-diffusion case (not shown) to be $1.6\times10^{-3}$
and $1.2\times10^{-3}$, respectively, and for the unknown-diffusion
case to be $1.7\times10^{-2}$ and $1.1\times10^{-2}$, respectively.
For reference, we also show exact-interpolated velocities in the right
subplots in figure \ref{fig:u_vel_analyt} and \ref{fig:v_vel_analyt},
i.e., velocities evaluated exactly along the 3000 locations along
the characteristics, then interpolated onto the regular grid used
for visualization.

Figure \ref{fig:vel_comvel_comp_uniform_unkn_diff} shows that throughout
the domain spanned by the characteristics emanating from the inital-condition
curve $\Gamma(t)$ (solid red line in the right panel of figure \ref{fig:Maslowe Vortices}),
there is generally close agreement between the reconstructed and the
exact velocity profiles. The domain of the velocity reconstruction
is limited by the weak mixing of the scalar field $c(\mathbf{x},t)$
in this steady flow. The initial condition curve in the right panel
of figure \ref{fig:Maslowe Vortices} only intersects the scalar contours
forming the three middle cells in the left panel of figure \ref{fig:Maslowe Vortices}.
Hence, only these cells appear in the reconstructed velocities on
the left in figures \ref{fig:u_vel_analyt} and \ref{fig:v_vel_analyt}.

\subsection{Example 2: Unsteady Navier\textendash Stokes velocity field}

\begin{figure}
	\centering
	\sidesubfloat[$t=490$]{\includegraphics[width=0.4\textwidth]{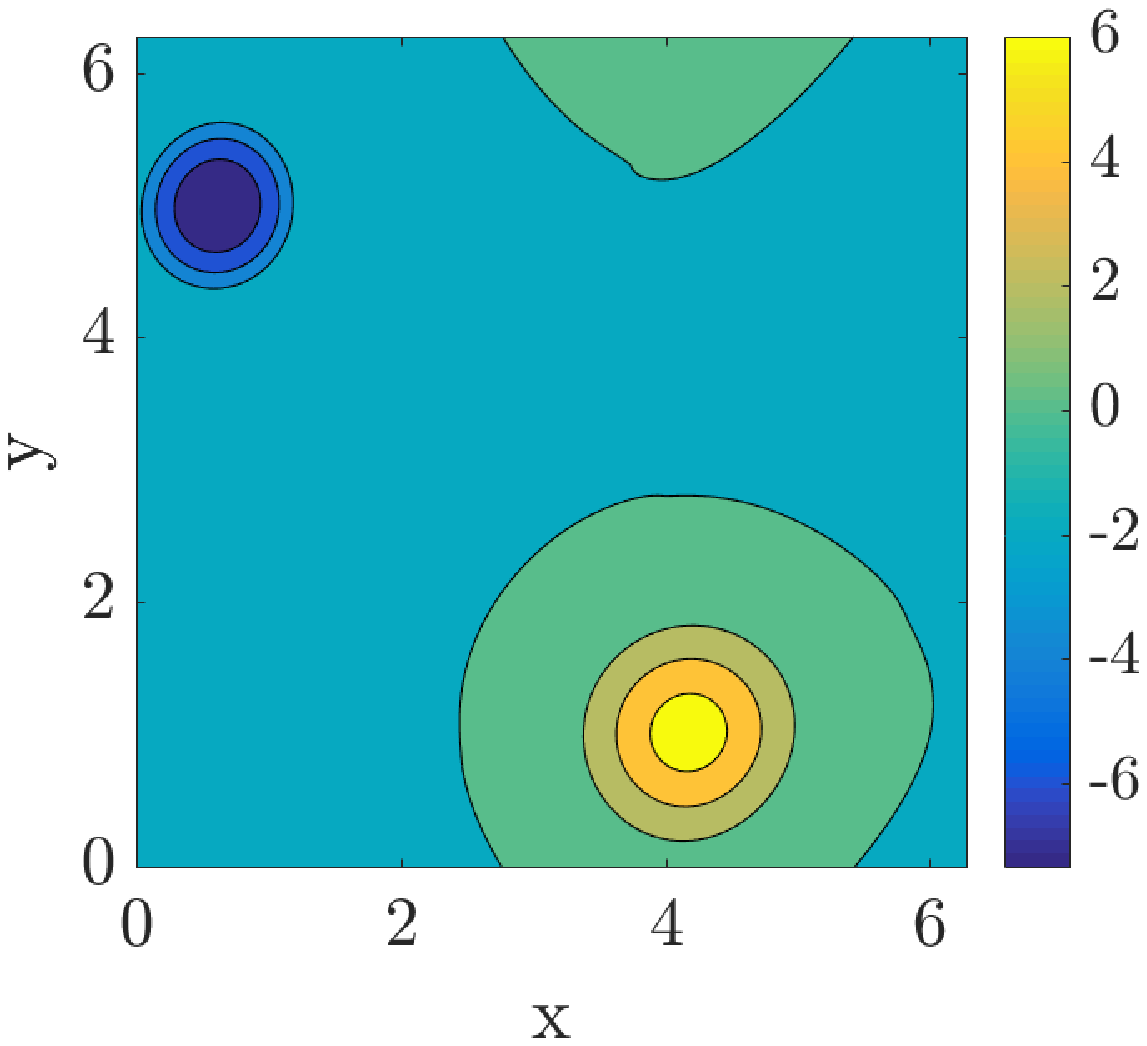}\label{fig:NS_Vorticity_t_0} }
	\sidesubfloat[$t=499.5$]{\includegraphics[width=0.4\textwidth]{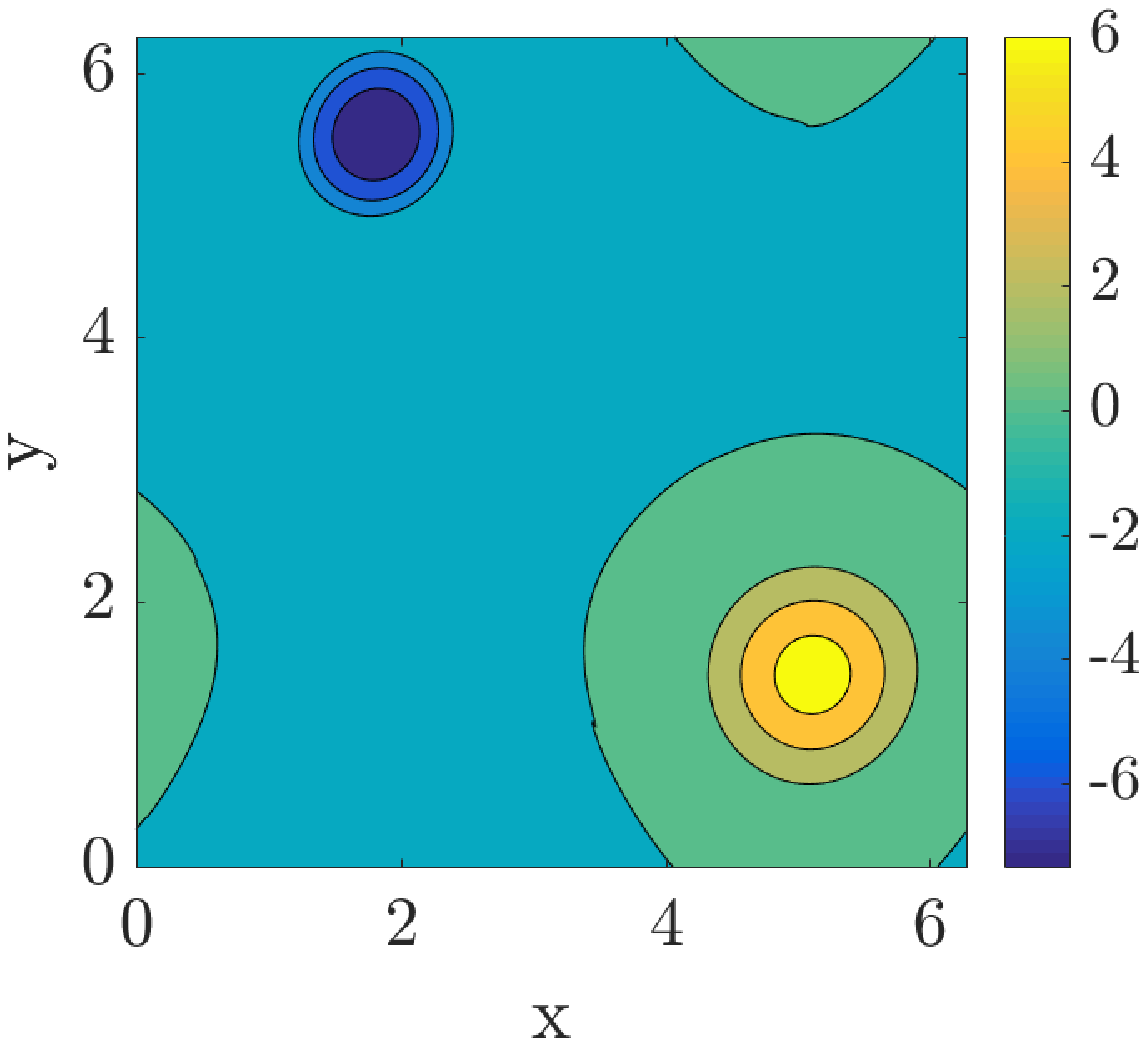} \label{fig:NS_Vorticity_t_10}}
	\caption {Evolution of vorticity field for Example 2: (a) $\omega(\mathbf{x},490)$ (b) $\omega(\mathbf{x},499.5)$.\label{fig:Navier Stokes Vorticity}}
\end{figure}

We now apply our velocity reconstruction procedure to an unsteady
flow obtained from the direct numerical solution of the unforced,
two-dimensional Navier\textendash Stokes equations. We solve the vorticity-streamfunction
formulation of the equations, as described, e.g., in \citet{melander1987axisymmetrization},
over the doubly periodic square domain $\left(x,y\right)\in[0,2\pi]\times[0,2\pi]$
with the non-dimensional kinematic viscosity $\nu=10^{-5}$. Hence,
the scalar-diffusion characterized by $\mu=0.01$ is faster than the
vorticity-diffusion, in contrast to the requirements of the SIV method
of \citet{dahm1992scalar}. We initialize the numerical simulation
with the velocity field 
\begin{align}
\begin{split}u(x,y,0)= & \hspace{0.15in}\sin(x)\cos(y)+\sin(3x)\cos(3y),\\
v(x,y,0)= & -\cos(x)\sin(y)-\cos(3x)\sin(3y).
\end{split}
\label{eq:initial velocity}
\end{align}
Eventually, two distinct regions of opposite vorticity with different
strengths emerge, subsequently evolving periodically within the domain,
as shown in figure \ref{fig:Navier Stokes Vorticity}. For longer
simulation times ($t\approx500$ here), these final structures exhibit
slow decay due to dissipation and weak non-linearity \citep{matthaeus1991decaying}.

\begin{figure}
	\centering
	\sidesubfloat[$t=490$]{\includegraphics[width=0.4\textwidth]{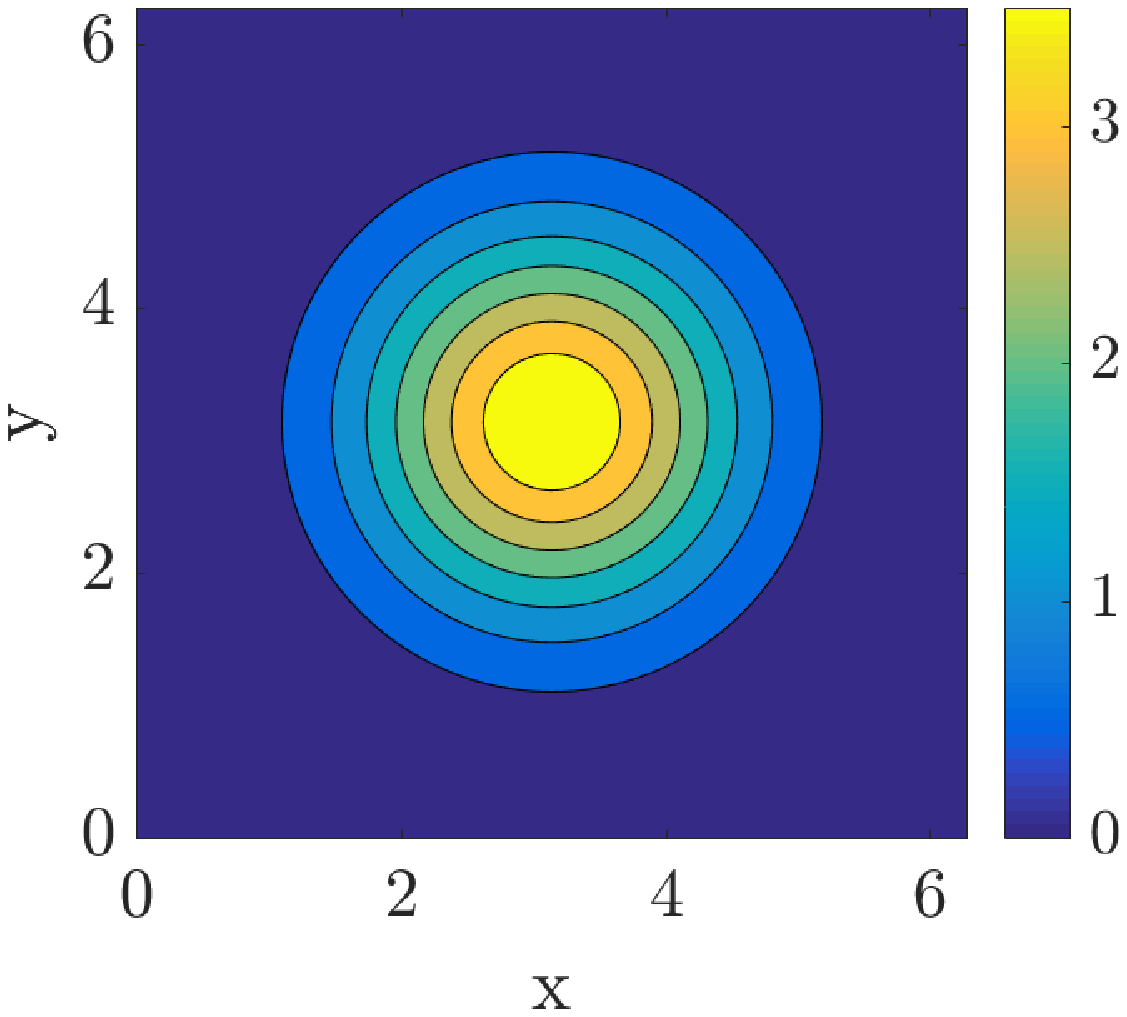}\label{fig:NSConc_0} }
	\sidesubfloat[$t=499.5$]{\includegraphics[width=0.4\textwidth]{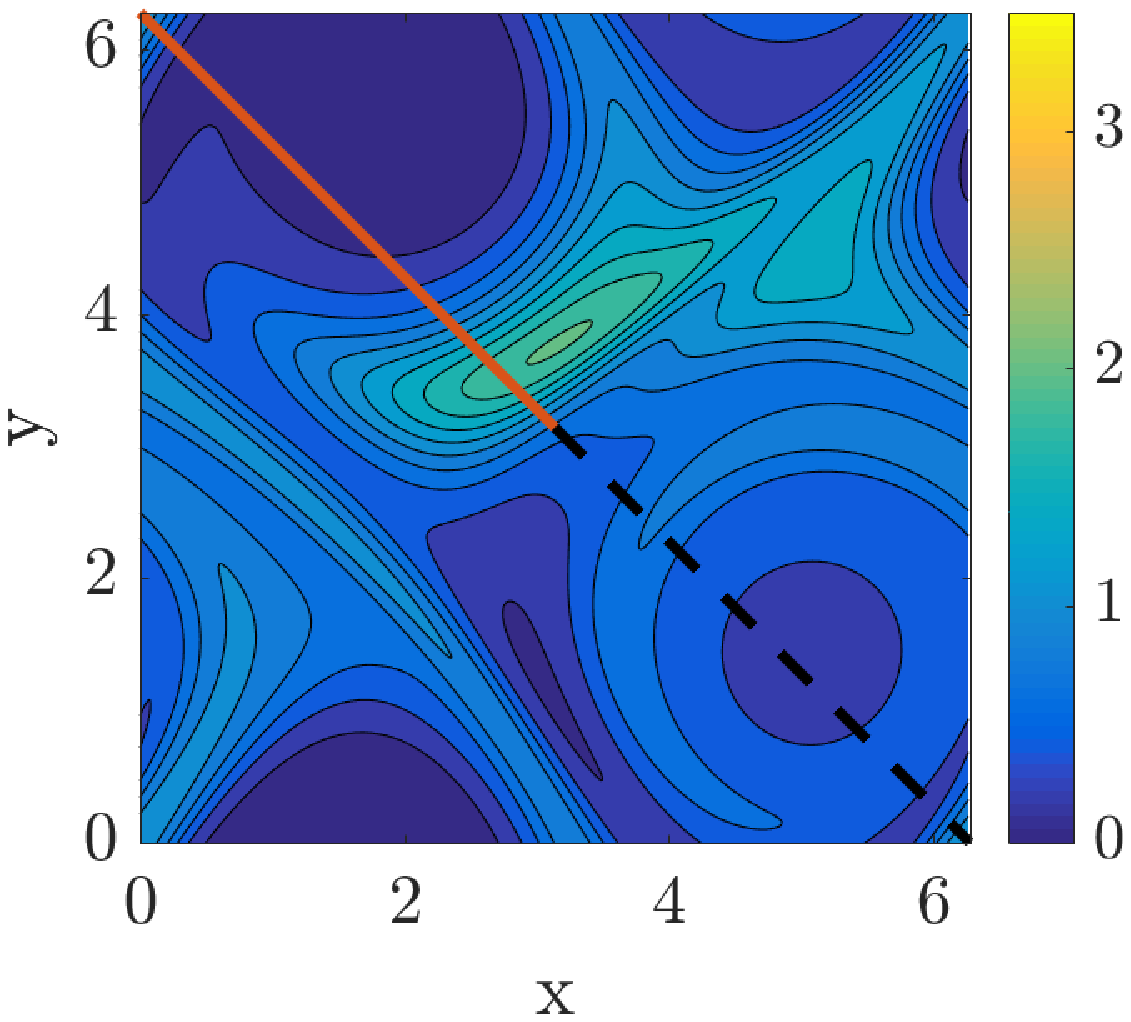}\label{fig:NSConc_10} }
	\caption {Evolution of the diffusive scalar $c(\mathbf{x},t)$ under the unforced, two-dimensional Navier\textendash Stokes flow initialized by the velocity field \eqref{eq:initial velocity} of Example 2: (a) $c(\mathbf{x},490)$ (b) $c(\mathbf{x},499.5)$. The solid red curve and the dashed black curve shown over the diffused concentration field are two different initial (non-characteristic) curves for the velocity reconstruction.\label{fig:NSConc}}
\end{figure}

Using the velocity field from $t_{0}=490$ onwards, we advect a scalar
field $c(\mathbf{x},t)$ with initial distribution 
\begin{equation}
c(\mathbf{x},t_{0})=4\exp[-0.5((x-\pi)^{2}+(y-\pi)^{2})],
\end{equation}
and with diffusivity $\mu=0.01$ (cf. figure \ref{fig:NSConc}). The
unsteady velocity field visibly leads to intense mixing compared to
Example 1. 

We have reconstructed the velocity field for various times between
$t_{0}=490$ and $t=499.5$ with similar outcomes in all cases. Here
we only show the result for $t=499.5$ in figures \ref{fig:NS_vel_line_1}
and \ref{fig:NS_vel_line_2} for the solid red and dashed black lines
of figure \ref{fig:NSConc} as initial curves, respectively.

\subsubsection{Solid red initial curve}

\begin{figure}
	\centering
	\sidesubfloat[u velocity]{\includegraphics[width=1\textwidth]{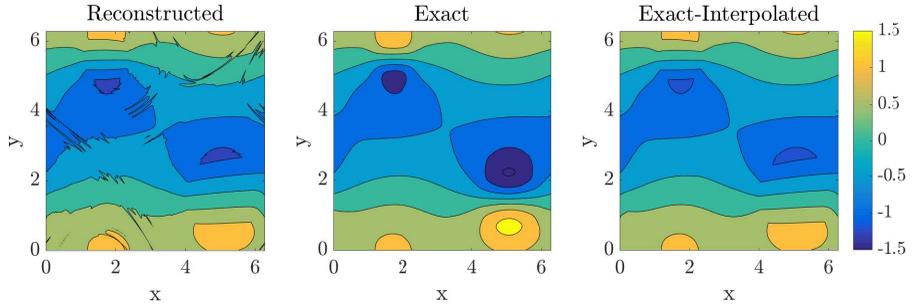}\label{fig:NS_u_vel_line_1}}\\
	\sidesubfloat[v velocity]{\includegraphics[width=1\textwidth]{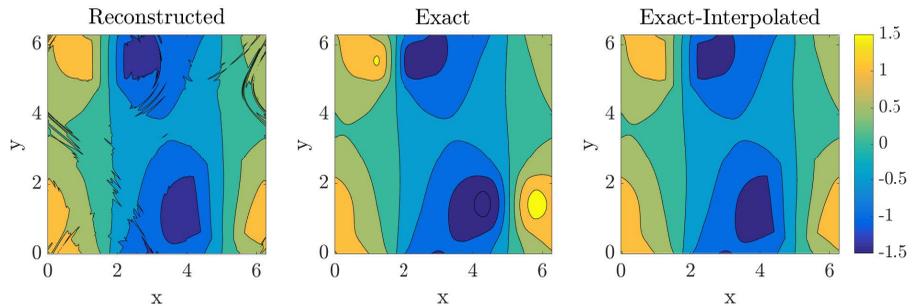} \label{fig:NS_v_vel_line_1}}
	\caption {Reconstructed, exact, and exact-reconstructed velocities for Example 2 (with the diffusivity, $\mu$, treated as unknown) at $t=499.5$, using the red-solid line from figure \ref{fig:NSConc_10} as the initial curve: (a) $u(\mathbf{x},499.5)$ (b) $v(\mathbf{x},499.5)$. \label{fig:NS_vel_line_1}}
\end{figure}

If the initial curve $\Gamma(t)$ for the characteristic ODEs \eqref{eq:characteristics}-\eqref{eq:ode1}
is the solid red curve in figure \ref{fig:NSConc}, then the reconstructed
velocity field captures most features of the true velocity field.
For the known diffusivity formulation, the reconstruction errors $E_{u}$
and $E_{v}$ are $1.9\times10^{-3}$ and $2.5\times10^{-3}$, respectively.
For the unknown diffusivity formulation, we have obtained $E_{u}=5.2\times10^{-2}$
and $E_{v}=5.9\times10^{-2}$. For the more erroneous latter case,
the overall qualitative accuracy is still compelling, as we show in
figure \ref{fig:NS_vel_line_1}. Most of the local inaccuracies arise
from interpolation errors, due to a lack of penetration of the external
scalar level curves into the cores of high-vorticity regions (cf.
the lower-right region of figure \ref{fig:NSConc_10}). This is confirmed
by a comparison of the exact-interpolated and the exact plots in figures
\ref{fig:NS_u_vel_line_1} and \ref{fig:NS_v_vel_line_1}. While the
spatial domain of the reconstructed velocity is limited by the presence
of scalar transport barriers, the numerical calculation is overall
accurate (cf. the reconstructed and exact-interpolated plots of figure
\ref{fig:NS_vel_line_1}). 

\subsubsection{Dashed\textendash black curve}

\begin{figure}
	\centering
	\sidesubfloat[u velocity]{\includegraphics[width=1\textwidth]{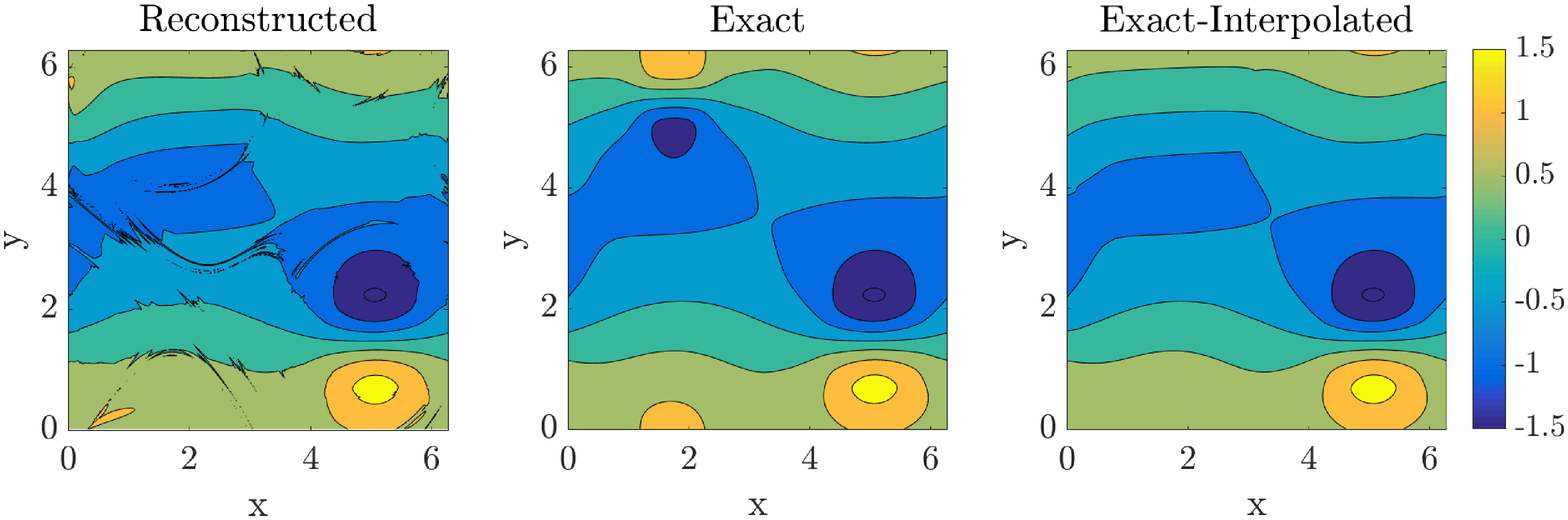}\label{fig:NS_u_vel_line_2}}\\
	\sidesubfloat[v velocity]{\includegraphics[width=1\textwidth]{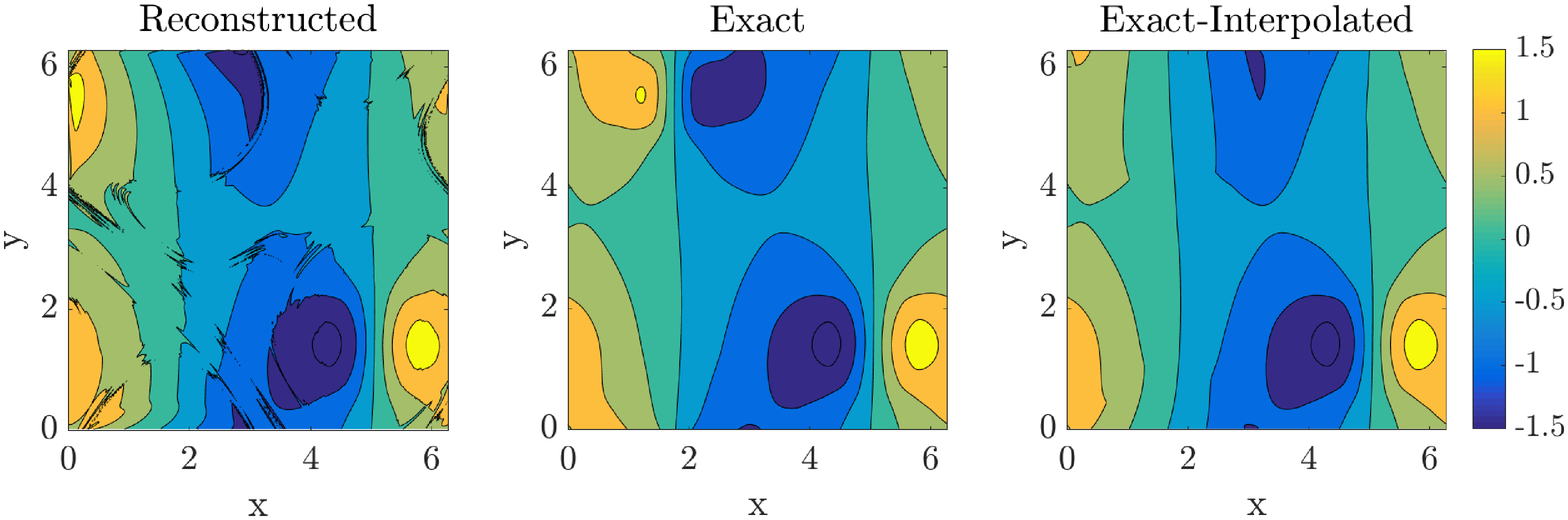} \label{fig:NS_v_vel_line_2}}
	\caption {Velocities for Example 2 (diffusivity, $\mu$, assumed unknown) at $t=499.5$ using black-dashed line from figure \ref{fig:NSConc_10} as initial curve: (a) $u(\mathbf{x},499.5)$ (b) $v(\mathbf{x},499.5)$. The plotting is done identical to figure \ref{fig:NS_vel_line_1}. \label{fig:NS_vel_line_2}}
\end{figure}

For this case (with the scalar diffusivity, $\mu$, assumed unknown),
the reconstructed velocity field recovers the positive high-vorticity
region (bottom right of figure \ref{fig:NS_Vorticity_t_10}), fully
capturing even the finer features (cf. the reconstructed vs. the exact
velocity field in figure \ref{fig:NS_vel_line_2}). However, the negative
vorticity region in the top left of figure \ref{fig:NS_Vorticity_t_10}
is not fully captured in the reconstructed velocity plots of figure
\ref{fig:NS_vel_line_2}. The error from this calculation is nevertheless
still low: $E_{u}=6.92\times10^{-2}$ and $E_{v}=6.23\times10^{-2}$.

\section{Oceanographic applications\label{sec:Oceanographic-applications}}

Oceanic velocities are key variables in understanding a number
of environmental processes that range from climate change and global
warming to transport of pollutants and various bio/geo/chemical tracers.
Obtaining reliable estimates for ocean currents with high accuracy
and resolution, however, remains a challenging task.

Oceanic velocities can be measured directly by current meters or Acoustic
Doppler Current Profiler (ADCP) instruments attached to ships, stationary
moorings, or towed or autonomous sampling platforms. Such estimates
provide vertical profiles or sections of velocities at select locations
or along select ship tracks but only at limited times and locations.
Ship-based ADCP sections, however, can provide the required initial
conditions for our proposed velocity reconstruction method along all
tracer contours intersecting the ship track.

Surface drifters also provide estimates of near-surface ocean currents.
These semi-Lagrangian instruments consist of a surface-drifting buoy
equipped with a GPS tracker and a subsurface drogue, or sail, underneath. Finite differencing consecutive GPS fixes with respect to the time interval between neighboring GPS transmissions
provides estimates for the underlying ocean currents at the depths of the drogue, typically 20m. The Global Surface Drifting Buoy Array maintained by
NOAA is part of their Global Ocean Observing System (http://www.aoml.noaa.gov/phod/gdp/objectives.php).
Currently, this drifter array has 1445 operational drifters that are
distributed throughout the World Oceans with at least 1 drifter per
each 5 x 5 deg bin at any given time. Although the 5 degree resolution
of the resulting velocity field obtained from these drifters via finite
differencing appears coarse, it would yield a much higher-resolution
global estimate of the near-surface oceanic velocities when used as
a boundary condition for the velocity inversion proposed here along
all observed tracer contours passing through drifters positions.

A third alternative for collecting ocean current data is indirect
remote sensing from either land-based or satellite-based radars. For
example, satellite-based estimates of sea surface height provide nearly global geostrophic
velocities at the ocean surface (\emph{cf.} https://www.aviso.altimetry.fr).
However, these velocity estimates only provide the geostrophic component of the total velocity field and only on spatial scales larger than meso-scale. Land-based high-frequency radars, on the other hand, can measure full surface velocity at a higher spatio-temporal resolution, but only over limited spatial domains. 

Satellite-based measurements of diffusive tracers\textendash such as sea surface temperature, sea surface salinity, or ocean color\textendash are also available. Applying the velocity reconstruction technique described in this paper to these tracer distributions, along with the available information about the localized velocity values from current meters, ADCPs, drifters, moorings or HF radars, would potentially allow for the estimation of the oceanic velocities in a much larger (possibly global) domain.

Clearly, the opportunities we have described above for ocean
current reconstruction come with their own challenges, including non-diffusive
decay of tracers, sporadic availability of the tracers in space and
time, presence of noise in data, lack of exact two-dimensionality and incompressibility, and
spatial dependence of the diffusivity. Some of these challenges we
have already addressed in sections \ref{subsec:Spatially-dependent-diffusivity}
and \ref{subsec:Shallow-water-approximation}, sensitivity to noise and undersampling will be explored in section \ref{sec:NoiseandUndersample}, and the rest we leave to
future work. As a proof of concept, we consider below a numerical
example in which we apply our proposed velocity inversion to a conservative
tracer in a barotropic ocean circulation model.

\subsection{Application to an ocean circulation model\label{subsec:Application-to-an}}

We seek to reconstruct velocities from a tracer distribution generated
numerically by the ocean circulation model MITgcm \citep{marshall1997finite}.
We keep the complexity of the model to the minimum and use the simplest
form of the advection-diffusion equation with a constant diffusivity
coefficient. In its linear limit, the model is configured to reproduce
the solution described by the \emph{Munk model} of the wind-driven
ocean circulation \citep{munk1950wind}, though the numerical model
allows for nonlinear features to develop, such as eddies and jets.

The simulation has been configured on a beta-plane with a rigid lid,
3.5 km spatial resolution, a temporal step of 200 seconds and a momentum
diffusion coefficient of 40 m$^{2}$s$^{-1}$. The fluid is unstratified
and initially is at rest, but is subject to a temporally constant
stress at the surface boundary that drives the gyre flows that develop.
Momentum input by the wind at the surface is removed by boundary friction
at the lateral boundaries. 

Consistent with the Munk model, we mimic
the time-mean-observed winds with an idealized structure that is zonally
uniform and exerts stress only in the zonal direction. The winds have
a cosine structure in the meridional direction, with westerly winds
at mid latitudes and easterly winds to the north and south. The net
vorticity input into the domain is zero, with the negative vorticity
input into the northern half of the domain balanced by the positive
vorticity in the south of the domain. A 2-gyre ocean circulation develops,
with intense western boundary currents mimicking the Gulf Stream and
Labrador currents of the Atlantic Ocean, and the Kuroshio and Oyashio
of the Pacific. The two gyres are separated by a meandering jet \textendash
an analogue of the Gulf Stream or Kuroshio Extension currents.

After an initial spin up of 300 days, a tracer with a constant meridional
gradient was released into the model, and its evolution over the subsequent
180 days was simulated numerically using an advection-diffusion scheme
with constant diffusivity coefficient $\mu =100$ m$^{2}$s$^{-1}$.
Three consecutive tracer snapshots 200 s apart were saved on day 180
and were used to numerically estimate matrices $\mathbf{A}$ and $\mathbf{b}$
in equation \eqref{eq:known_diff_form}, for use with the velocity inversion
equation \eqref{eq:ode1}. All tracer derivatives were computed using finite differences, in a manner similar to what is feasible in ocean applications involving remotely sensed velocity and tracer data.

\begin{figure}
	\centering
	\includegraphics[width=1\textwidth]{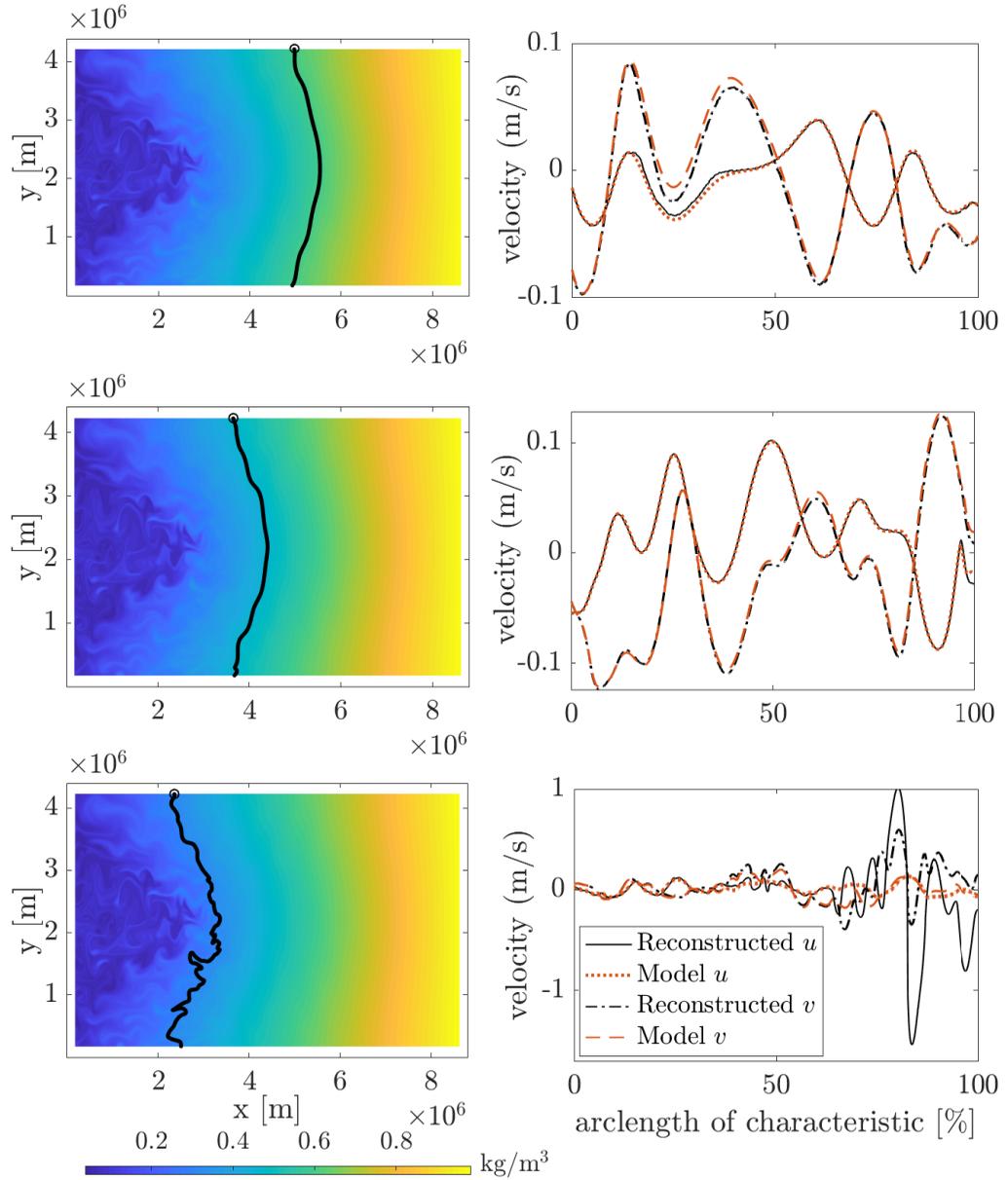} 
	\caption {Right: Comparison between true and reconstructed u and v velocity components along the 3 representative tracer iso-contours in an idealized model of the North Atlantic Ocean. Left: Tracer distribution with the corresponding iso-contours.}
	\label{fig:Unsteady2Gyre_run7}
\end{figure}

We have applied the velocity inversion method with known-scalar diffusivity (eqs. \eqref{eq:ode1}-\eqref{eq:known_diff_form}) to
reconstruct velocities along several tracer iso-contours covering different parts of the model domain. We show a comparison between the reconstructed velocity and the ground truth (actual model velocity on day 480) in figure \ref{fig:Unsteady2Gyre_run7} for three representative tracer contours. In this calculation, we assume that the true velocity\textemdash our initial condition for integrating eq. \eqref{eq:ode1}\textemdash is known at the northernmost point of the contour, then perform the integration going southward along the entire length
of the contour. The reconstructed velocities provide useful information
for oceanographic studies in roughly the right 2/3 of the model domain,
far enough from the western boundary where the eddy activity is most
vigorous and the tracer contours are most filamented. Close to the
western boundary, in the left 1/3 of the domain, the reconstructed
velocity diverges from the true model velocity due to numerical errors
arising in the computation of $\mathbf{A}$ and $\mathbf{b}$ and
in the integration of the matrix equation \eqref{eq:ode1} along the
tracer contours. The reconstructed velocity is still useful along
the initial segment of the iso-contour before the error accumulation
causes significant mismatch. The reconstruction performs better if we
integrate from the equator going poleward to the north and south (thus
decreasing the length of the tracer contour two times), or if the
true velocity is known at more than one point along the tracer contour.
In that case, we can apply the velocity inversion to the segments
of the contour instead of the entire contour.

\begin{figure}
	\centering
	\includegraphics[width=1\textwidth]{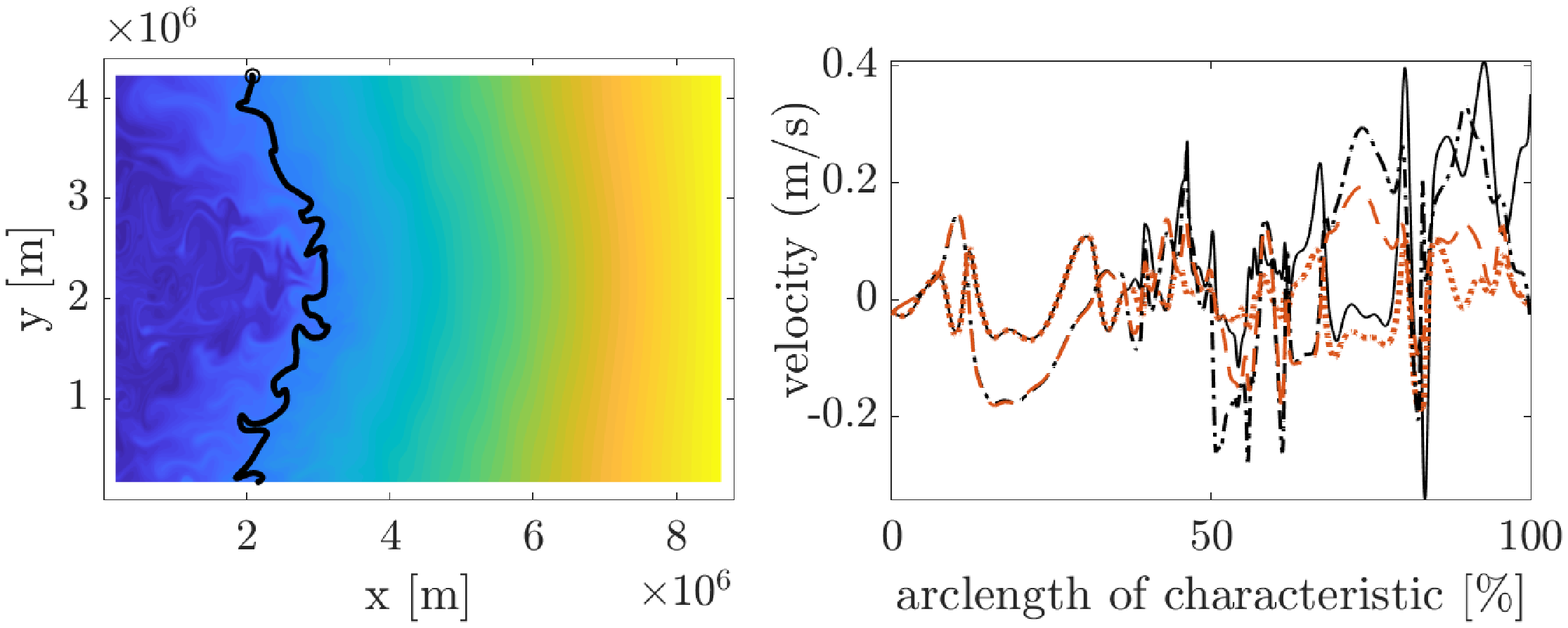} 
	\caption {Same as in figure \ref{fig:Unsteady2Gyre_run7} but using a spatially-averaged tracer field.}
	\label{fig:Averaged_west}
\end{figure}

{In the case when the spatio-temporal resolution of the scalar field is significantly higher than the dominant length and/or time-scales of the flow, the reconstruction can be improved by smoothing the tracer field in space and/or time to suppress noise arising during the calculation of tracer derivatives. In our example, applying a spatial smoothing, specifically, the two-dimensional convolution filter `conv2' in MATLAB with a constant-coefficient 6x6 square kernel (equivalent to a two dimensional running average over $\sim$21 km$^2$), allows for reasonably accurate velocity reconstruction further into the left 1/3 of the domain (figure \ref{fig:Averaged_west}).}

{\subsection{The effect of resolution and uncertainty in the data}\label{sec:NoiseandUndersample}
While our analytic velocity reconstruction is, in principle, exact under conditions described in section \ref{sec:Formulation}, its practical implementations will have to address sparsity and uncertainty in the available observational data. A general analysis of these issues with real ocean data is beyond the scope of this paper, but we provide an initial illustration of the sensitivities to noise and resolution for the oceanographic model we have studied.}

\begin{figure}
	\centering
	\subfloat{\includegraphics[width=0.49\textwidth]{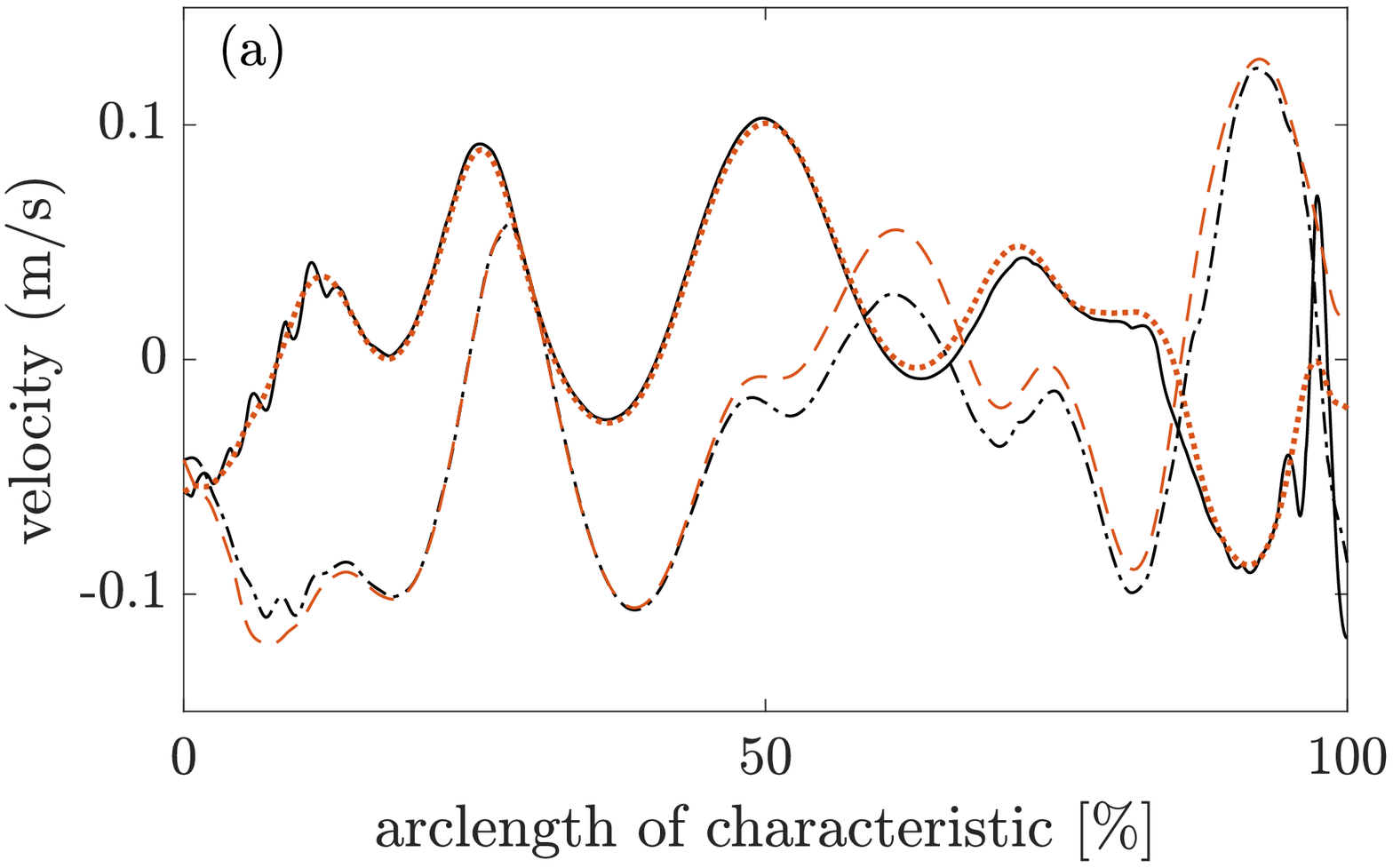}\label{fig:figure9a} }
   \subfloat{\includegraphics[width=0.49\textwidth]{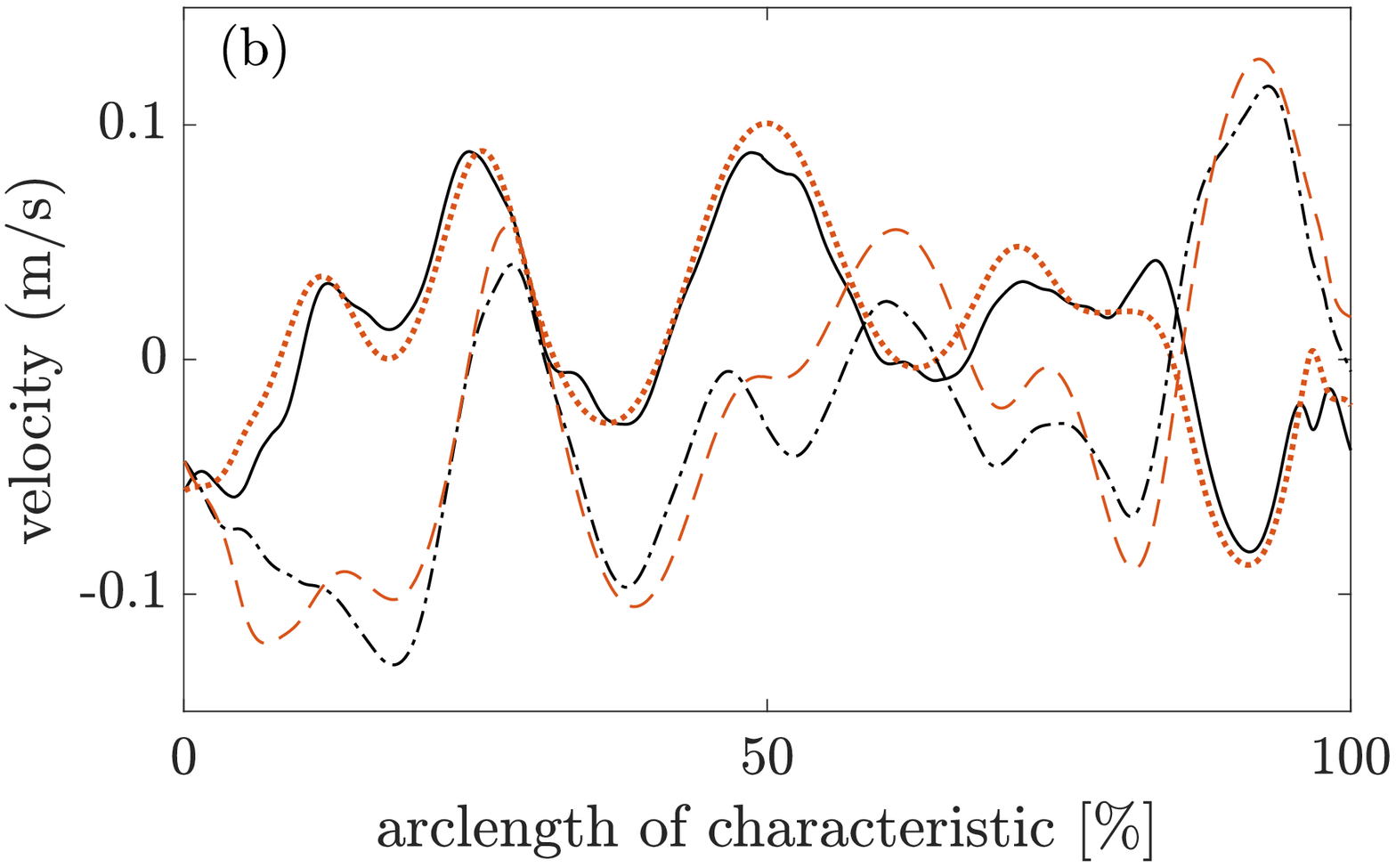} \label{fig:figure9b}}\\
	\subfloat{\includegraphics[width=0.49\textwidth]{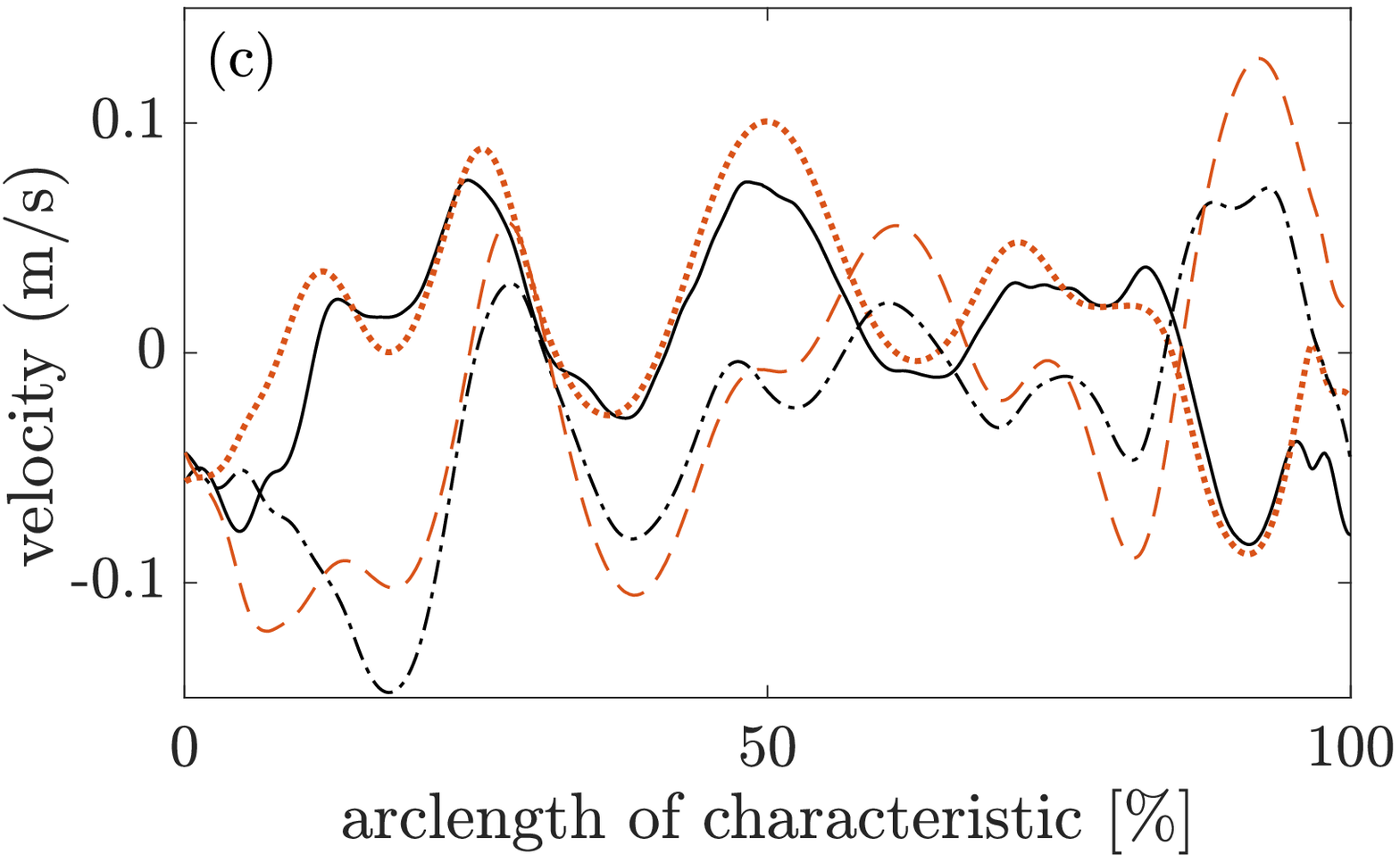}\label{fig:figure9c} }	\subfloat{\includegraphics[width=0.49\textwidth]{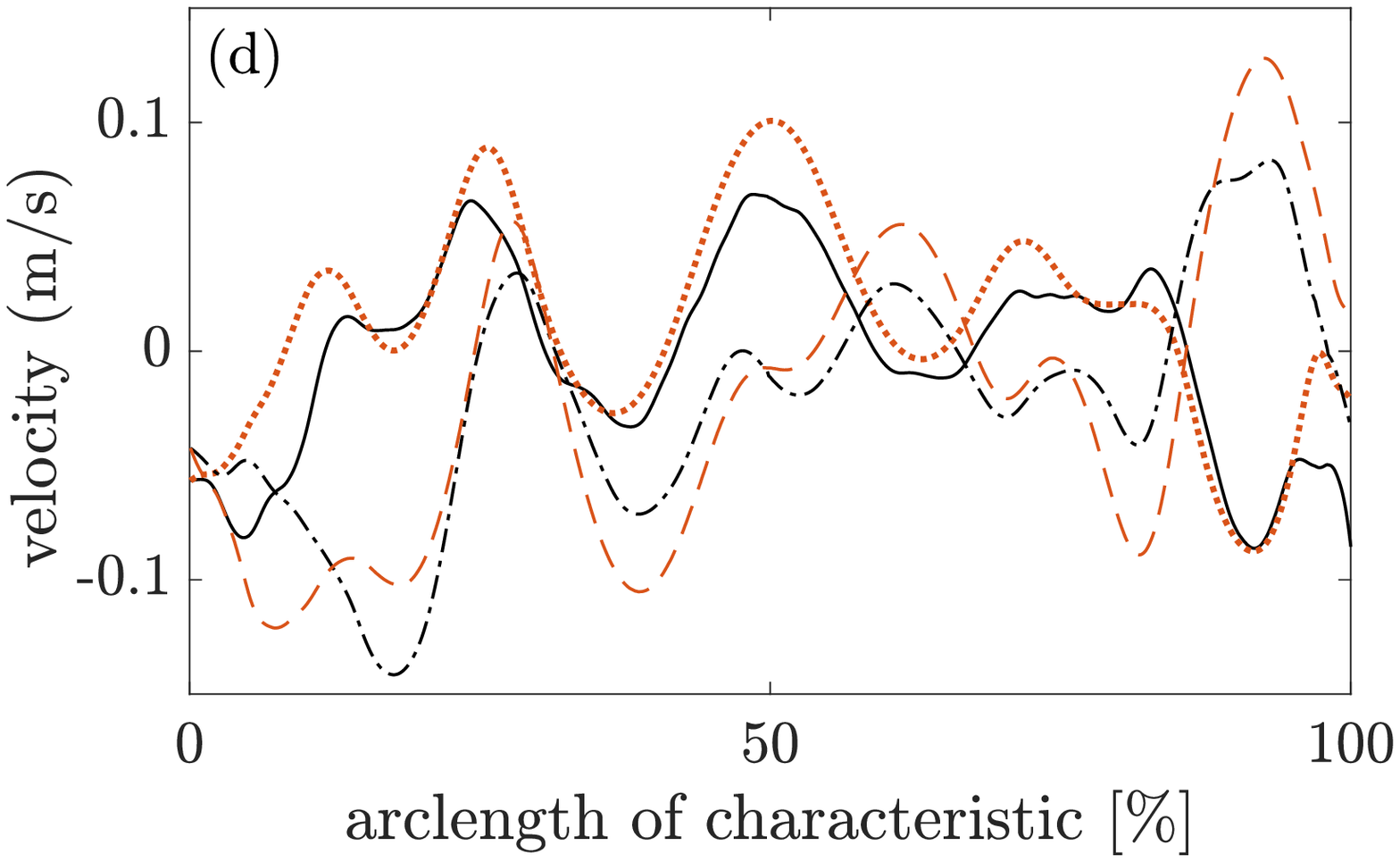}\label{fig:figure9d} }
	\caption {Velocity reconstruction along the contour shown in middle panel of figure \ref{fig:Unsteady2Gyre_run7} (same legend for curves), but with sparse/ noisy tracer data. (a) 20 times sparse tracer data (resolution changed from 3.5 km to 70 km) in each direction; (b) zero mean $\tanh(\text{Gaussian})$ noise with Gaussian standard deviation 0.01 and amplitude 10\% of the local temporal mean of the tracer added to the tracer data; (c) as (b) but with 20\% noise amplitude; (d) Combination of (a) and (c).}
	\label{fig:figure9}
\end{figure}

\begin{figure}
	\centering
	\subfloat{\includegraphics[width=0.49\textwidth]{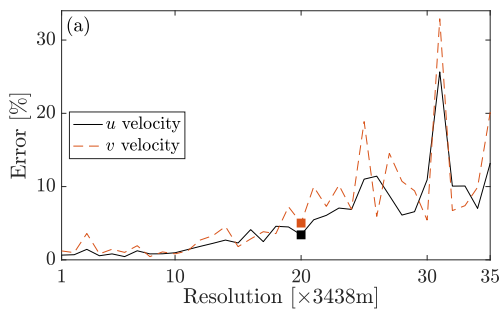}}
	\caption {Contour-averaged error as a percentage of the velocity range along the contour  shown in middle panel of figure \ref{fig:Unsteady2Gyre_run7} as a function of spatial resolution. 
		The square markers represent the case in figure \ref{fig:figure9a}.}
    \label{fig:figure10}
\end{figure}

{We illustrate the outcome of velocity reconstruction from under-resolved tracer fields in figures \ref{fig:figure9a} and \ref{fig:figure10}, where we sub-sampled the full model tracer field over an increasingly sparse grid. The sub-sampled field was then interpolated to its original grid using MATLAB's build-in cubic interpolator, and the velocity inversion was applied along the same level contour of the tracer, as in the middle panel of figure \ref{fig:Unsteady2Gyre_run7}. An example of the velocity reconstruction using an under-resolved tracer field is shown in figure \ref{fig:figure9a}.} 

{Our analysis suggest that the inversion is reasonably insensitive to under-sampling as the contour-averaged reconstructed velocity error (expressed as a percentage of the range of respective velocity along the iso-contour) in both zonal and meridional direction (figure \ref{fig:figure10}) is less than $5\%$ for grids with spacing up to about 70 km which is 20 times the original grid spacing. The critical grid spacing is likely dependent on the kinetic energy spectrum and the dominant length scales of the flow.}

{Since real data is noisy, we have also investigated the applicability of the velocity inversion to model fields with increasing noise levels. We add noise pointwise,
\begin{equation}\label{eq:noise}
N=A\tanh{(\mathcal{N}(0,\sigma))},
\end{equation} 
where $A$ is the noise amplitude and $\mathcal{N}(0,\sigma)$ is a zero mean Gaussian random variable with standard deviation $\sigma$ (generated using MATLAB's `randn' function), to the model tracer field. Unlike a pure Gaussian signal, \eqref{eq:noise} allows for defining a maximal noise amplitude normally guaranteed for a remote-sensing instrument. As a first step in our reconstruction algorithm, we use a linear regression model using the 37 time slices (about 2 hours) centered around the current time to temporally filter the input scalar-field data. We then smooth the noisy tracer concentration fields in space using MATLAB's `conv2' filter with constant coefficient square kernels, both before and after calculating a scalar derivative required in the reconstruction (\textit{cf.} equation \eqref{eq:known_diff_form}). For purely spatial derivatives of any order, a 17$\times$17 kernel is used in all the examples presented below, which correspond to about 58 km. The size of spatial filter for terms involving temporal derivatives is optimised for each case.}

{Figure \ref{fig:figure9b} shows the velocity reconstruction along the tracer contour in the middle panel of \ref{fig:Unsteady2Gyre_run7} but using the noisy scalar data with noise amplitude  $A$ equal to $10\%$ of the local temporal mean concentration across the 37 time slices (equivalent to a 2-hour period, which is effectively the observation time for this example) and setting a standard deviation $\sigma$ equal to $0.01$ in the definition \eqref{eq:noise}. The shape of the iso-contour remains almost the same as that without the noise. The kernel size for spatial filtering of terms involving the temporal derivatives is 30$\times$30 or, equivalently, 103 km. For these parameters, we have found the contour-averaged errors relative to the model velocities in the zonal and meridional direction to be 5.6\% and 7.8\%.}

{In the remote sensing of scalars, such as sea-surface temperature through satellite microwave radiometers, noise levels up to 18\% may be present \citep{gentemann2010accuracy}. Aerial measurements of dye from multispectral and hyperspectral cameras have similar noise levels (about 20\%) for Rhodamine WT concentrations of about 20ppb \citep{hally2015surfzone,clark2014aerial}. Dye releases of this magnitude on spatio-temporal scales of several km and several hours have been successfully carried out both in the surf-zone \citep{hally2015surfzone,clark2014aerial} and in the coastal ocean \citep{rypina2016investigating}. Figure \ref{fig:figure9c} shows the reconstructed velocity with 20\% noise amplitude and the rest of the parameters being the same as figure \ref{fig:figure9b}. The kernel size for spatial filtering of terms involving the temporal derivatives is 45$\times$45 (or 155 km). The contour-averaged errors in zonal and meridional velocity for this case are 9\% and 10.8\%. The joint effect of 20\% amplitude noise and under-sampling (20 times less resolution) is shown in figure \ref{fig:figure9d}. The contour-averaged errors in the zonal and meridional velocity are 10\% and 10.5\% for this case.}

{The presence of second and third order derivatives in equation \eqref{eq:known_diff_form} would usually make our reconstruction method susceptible to noise in the data. In the example presented above, however, we find that $\mathcal{O}(\frac{\partial^2c}{\partial t\partial x},\frac{\partial^2c}{\partial t\partial y})\gg \mathcal{O}(\mu\frac{\partial{\nabla^{2}c}}{\partial x},\mu\frac{\partial{\nabla^{2}c}}{\partial y})$ holds throughout the domain, so the right-hand side of equation \eqref{eq:known_diff_form} is dominated by the second\textemdash rather than the third\textemdash derivative terms. Recall that for a characteristic velocity, $U$, and characteristic length scale, $L$, the P\'eclet number for the scalar transport equation \eqref{eq:Transport Equation} is defined as
\begin{equation}
Pe=\frac{UL}{\mu}.
\end{equation}
Using the empirical relation between length scale $L$ [cm] and diffusivity $\mu$ [cm$^2$/s] obtained through synthesis from various dye release experiments in the ocean \citep{okubo1971oceanic}, we can let $\mu=0.0103 L^{1.15}$. Therefore, for typical velocity values, $U=10$--100 cm/s, the P\'eclet number estimated for length scales, $L=0.5$--10$^4$ km ranges from 61 to 2706. Evolution of tracers in flows with large P\'eclet numbers is advection-dominated, rendering the terms involving the third derivatives in \eqref{eq:known_diff_form} small. Thus, in our velocity reconstruction procedure, the sensitivity to noise would enter largely through the second derivatives in eqs. \eqref{eq:known_diff_form} or \eqref{eq:varying_diff_forcing}. Therefore, the most effort for noise suppression in the purely spatial derivatives via smoothing or filtering should be directed towards these terms. }

{In the noisy scalar examples presented in figures \ref{fig:figure9b} to \ref{fig:figure9d}, the major source of error is the first-order temporal scalar gradient, $\partial c/\partial t$. Using noisy scalar data to evaluate purely spatial derivative terms of all orders in equation \eqref{eq:known_diff_form} and using model scalar data for the temporal derivatives, we have found that an accurate velocity reconstruction (where contour average velocity errors remains less than about 10\%) can be obtained for much higher noise levels (both $A$ and $\sigma$ in equation \eqref{eq:noise}). Hence, the higher-order derivatives in the formulation of our velocity reconstruction method do not necessarily pose a challenge for accurate velocity reconstruction from noisy scalar data.}

\vskip 1cm 
{In summary, our numerical results indicate that the velocity reconstruction proposed here is of value to physical oceanography when applied to tracer fields with spatial resolution of several km, temporal resolution of minutes to hours, and noise levels of up to 20\% of the local temporal mean and a standard deviation of 0.01 for the noise model represented by equation \eqref{eq:noise}. Such resolution can be achieved for airplane-mounted radars but is currently beyond the capabilities of the existing satellites that have resolution from tens to hundreds of km spatially and from days to weeks temporally.}
	
{Even though our idealized model was able to reproduce several of the major qualitative features of the real ocean circulation in the Atlantic and Pacific ocean, a more realistic baroclinic free-surface model configuration is required in order to better simulate real oceanic conditions. We also note that because sea surface temperature and salinity are influenced by vertical mixing, air-sea interaction, and precipitation/evaporation/river run off, these two tracer fields are only purely diffusive on short temporal scales.  Additional challenges will arise in real ocean applications that are associated with small but non-zero horizontal divergence of surface velocity, which was exactly zero in our numerical model due to the rigid lid approximation, the space-dependent diffusivity, and the presence of higher-level non-Gaussian noise in data.}

\section {Conclusion}

It is well established that unique velocity reconstruction from an
observed scalar field is an underdetermined problem \citep{dahm1992scalar,kelly1989inverse}.
We have pointed out here, however, that if the velocity at a point
is known, then the velocity field all along the scalar level curve
emanating from that point can be uniquely expressed using sufficiently
well-resolved observations of the scalar. In a well-mixing flow, therefore,
velocities can in principle be reconstructed on global domains from
highly localized velocity measurements combined with global imaging
of a scalar field.

While our procedure also works with unknown scalar diffusivities,
we have found in our examples the reconstruction errors $E_{u}$ and
$E_{v}$ to be about an order of magnitude smaller when the diffusivity
is \textit{a priori} known. This is due to the numerical errors in
the evaluation of higher scalar derivatives involved in the unknown
diffusivity formulation (\emph{cf.} equation \eqref{eq:unknown_diff_form}
vs. \eqref{eq:known_diff_form}). To further improve accuracy for
the latter formulation, once the velocity field is recovered, a pointwise
estimate for the scalar diffusivity could be obtained at each point
in the domain spanned by the scalar level curves emanating from an
initial curve. A subsequent statistical analysis could then establish
an acceptable estimate for $\mu$ from these pointwise values, which
would in turn enable use of the known-diffusivity formulation of our
approach. Depending on accuracy requirements, this modification could
even be turned into an iterative procedure, involving a gradual, simultaneous
refinement of both the velocity field and the diffusivity estimate.

Extensions of our results to three-dimensional flows may be possible
but remain mathematically challenging because of the lack of a general
recipe for solving multi-dimensional systems of linear partial differential
equations. An extension to general compressible flows appears beyond reach because incompressibility or specific form of shallow water equations secures the common set of characteristic curves critical to our analytic solution strategy. A case with temporally constant but spatially varying density of the fluid in 2D will have a similar formulation as the shallow water case.

{The approach proposed here has the potential to extend spatially localized
ocean velocity measurements along contours of satellite-observed tracer
fields, such as temperature, salinity and ocean color (phytoplankton)
measurements. In Section \ref{sec:Oceanographic-applications}, we have provided a first proof of concept by reconstructing larger-scale velocities generated by a global circulation model, with the  tracer-field output of the same model serving as observational  input to our procedure. We have also demonstrated the robustness of this reconstruction under sparsification of  the observed  scalar field and the addition of noise that models observational inaccuracies.} More work is required to investigate applicability of this method to real ocean data. Further work can build on the extension of our theory we have
given in sections \ref{subsec:Spatially-dependent-diffusivity}-\ref{subsec:Shallow-water-approximation}
for spatially dependent diffusivities and 2D-divergent but mass-conserving
shallow water velocities.

\vskip 1 cm

We acknowledge useful conversations with Markus Holzner and Larry
Pratt, as well as partial funding from the Turbulent Superstructures
priority program of the German National Science Foundation (DFG),
and from the NASA grant \#NNX14AH29G to IR. 

\bibliography{Main documentarxiv}
\bibliographystyle{jfm}
\end{document}